\documentclass{myelsart} 
\usepackage{graphics}
\usepackage{amssymb} 
\usepackage{amsmath}
\usepackage{overcite} 

\begin{document}

\begin{frontmatter}

\title{Loop-closure events during protein folding: Rationalizing the shape of $\Phi$-value distributions}

\author{Thomas R.\ Weikl}
\address{Max-Planck-Institut f\"ur Kolloid- und Grenzfl\"achenforschung, 14424 Potsdam, Germany}

\begin{abstract}
In the past years, the folding kinetics of many small single-domain proteins has been characterized by mutational $\Phi$-value analysis. In this article, a simple, essentially parameter-free model is introduced which derives folding routes from native structures by minimizing the entropic loop-closure cost during folding. The model predicts characteristic folding sequences of structural elements such as helices and $\beta$-strand pairings. Based on few simple rules, the kinetic impact of these structural elements is estimated from the routes and compared to average experimental $\Phi$-values for the helices and strands of 15 small, well-characterized proteins. The comparison leads on average to a correlation coefficient of 0.62 for all proteins with polarized $\Phi$-value distributions, and 0.74 if distributions with negative average $\Phi$-values are excluded. The diffuse $\Phi$-value distributions of the remaining proteins are reproduced correctly. The model shows that $\Phi$-value distributions, averaged over secondary structural elements, can often be traced back to entropic loop-closure events, but also indicates energetic preferences in the case of a few proteins governed by parallel folding processes. 
\end{abstract}

\end{frontmatter}

\section{Introduction}

Small single-domain proteins with less than 100 amino acids typically are two-state folders \cite{jackson98,fersht99,grantcharova01}. These proteins fold from the denatured to the native state without populating experimentally detectable intermediate states \cite{fersht99}. In recent years, the folding kinetics of many two-state proteins has been characterized by mutational $\Phi$-value analysis. A $\Phi$-value is a measure for the impact of a mutation on the folding kinetics, defined as 
\begin{equation}
\Phi=- \frac{R T\ln k'/k}{\Delta G' - \Delta G} \label{phi}
\end{equation}
where $k$ and $\Delta G$ are the folding rate and stability of the wildtype protein, and $k'$ and $\Delta G'$ are the corresponding quantities of the mutant \cite{matouschek89,fersht99}. For various two-state proteins, detailed $\Phi$-value distributions have been obtained by considering many single-residue mutations throughout the protein chains.  A central question is why some proteins have polarized $\Phi$-value distributions, while others have diffuse distributions. In a polarized distribution, the $\Phi$-values for mutations in some of the secondary structural elements of the protein are significantly larger than the values in other secondary elements. In a diffuse distribution, the average $\Phi$-values for the secondary elements of the protein are rather similar.

Several results seem to indicate that the folding kinetics of two-state proteins is dominated by their native-state topology \cite{baker00}. Most importantly, the folding times of two-state folders have been found to correlate with the relative contact order (CO) of their native structures \cite{plaxco98,plaxco00,kuznetsov04,dinner01,ivankov03}. The relative CO of a protein is the average contact order, or `localness', $|i-j|$ of all native contacts $(i,j)$, divided by the chain length of the protein.  Here, $i$ and $j$ indicate the sequence positions of two residues in contact. The correlation holds for folding times over 6 orders of magnitude, from microseconds for $\alpha$-helical proteins with low relative CO to seconds for $\beta$-sheet-containing proteins with high relative CO. Comparable correlations with folding times have also been found for other measures of native-state topology \cite{gromiha01,makarov02,micheletti02,micheletti03,gong03,nolting03}. 
 
However, reproducing detailed $\Phi$-value distributions in theoretical models which are based on native-state topology has been proven to be difficult. Several theoretical models derive folding routes or $\Phi$-values from native structures. \cite{alm99,munoz99,galzitskaya99,guerois00,ivankov01,alm02,bruscolini02,karanicolas03,clementi00,hoang00,li01,karanicolas02,ding02,clementi03,brown04,portman01,kameda03,garbuzynskiy04}.
Some of these models assume that the amino acid residues can be in either of two states, native-like folded or unfolded \cite{alm99,munoz99,galzitskaya99,guerois00,ivankov01,alm02,bruscolini02,karanicolas03}. In this respect, the models are similar to the Zimm-Bragg model  for helix-coil transitions where residues can either be in a helix or in a coil state  \cite{zimm59}, or to Ising models where particles can either have spin up or down. Other models use explicit chain representations of the proteins and simplified Go-type potential energies which impose the native structure by postulating favorable interaction energies only between pairs of residues that are in contact in the native structure \cite{clementi00,hoang00,li01,karanicolas02,ding02,clementi03,brown04}. 
The unfolding kinetics of proteins has also been considered in Molecular Dynamics simulations with all-atom models \cite{li96,lazaridis97,gsponer01,dejong02,paci02,morra03}. Some of these models have been used to calculated $\Phi$-values either for a single protein or a small number of proteins \cite{alm99,munoz99,galzitskaya99,clementi00,guerois00,li01,karanicolas02,clementi03,brown04}.  A systematic comparison for a set of 19 proteins has been performed by Alm et al.\ with an Ising-like model \cite{alm02}. For more than half of the proteins, Alm et al.\ obtain correlation coefficients $r$ from 0.41 to 0.88 between theoretical and experimental $\Phi$-values, and for 14 of the 19 proteins, the theoretical $\Phi$-values were better than random permutations of the experimental values.  Kameda \cite{kameda03} has considered a Gaussian chain model with a Go-type interaction potential and obtains positive correlation coefficients $r$ between 0.12 and 0.65 for 7 out of 12 proteins. More recently, Garbuzynskiy et al.\ \cite{garbuzynskiy04} have reproduced the $\Phi$-value distributions of 17 proteins with an average correlation coefficient of 0.54. 

 The model presented here focuses on {\em average} $\Phi$-values for the secondary structural elements of a protein. The starting point of the model are native contact maps. The native contact map of a protein is a matrix in which element $(i,j)$ equals 1 if the two residues $i$ and $j$ are in contact in the native structure, and 0 otherwise. The contacts in the native contact map of a protein typically are arranged in clusters. These contact clusters correspond to structural elements such as $\alpha$-helices and $\beta$-strand pairings. 

\begin{figure}
\begin{center}
\resizebox{0.8\linewidth}{!}{\includegraphics{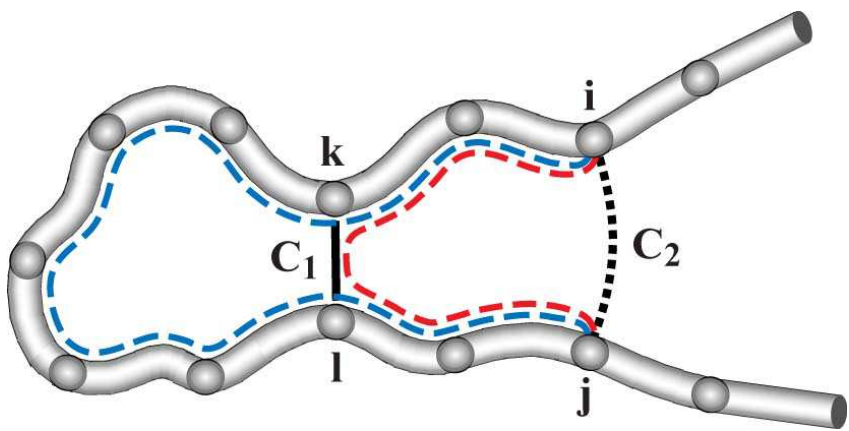}}
\end{center}
\vspace{1cm}
\caption{The effective contact order (ECO) for the contact $C_2$ is the length of the shortest path between the two residues $i$ and $j$ forming the contact. The `steps' in this shortest-path problem are either covalent bonds between adjacent residues, or noncovalent contacts formed previously in the folding process such as the contact $C_1$. In this example, the ECO for the contact $C_2$ is 5, since the shortest path (shown in red) involves two steps from $i$ to $k$, one step for the contact $C_1$ between $k$ and $l$, and two steps from $l$ to $j$. The ECO is a measure for the length of the loop which has to be closed to form the contact. In contrast, the contact order (CO) of $C_2$ is the sequence separation $|i-j|$ between the two residues, the number of residues along the blue path between $i$ and $j$. In this example, the CO for the contact $C_2$ is 10.
\label{eco}}
\end{figure}

In the first step, the model derives folding routes from native contact maps. In this step, the model considers all sequences in which the contact clusters, or structural elements, can be formed. The key assumption of the model is that the dominant folding routes can be identified as those sequences of events which minimize the loop-closure cost and, hence, the entropic barriers during folding. The loop-closure cost of a folding sequence simply is defined as the sum of loop lengths for forming the clusters along that sequence. These loops lengths are estimated via the graph-theoretical concept of effective contact order ECO \cite{fiebig93,dill93} (see Fig.~\ref{eco}). The ECOs and, thus, the loop-closure cost for forming nonlocal structural elements typically can be reduced by the previous formation of other, more local structural elements.  

In the second step, the model estimates the kinetic impact of contact clusters and secondary structural elements from the folding routes. In the model, the kinetic impact of a contact cluster depends on how often the cluster appears on the folding routes to (other) nonlocal clusters, and on the ECOs of the cluster. The kinetic impact derived from the folding routes is compared to average experimental $\Phi$-values for the secondary structural elements. To test the model systematically, 15 proteins are considered which (i) are small in the sense that they have less than 10 contact clusters, and which (ii) are well-characterized in the sense that $\Phi$-values for at least 10 residue positions are available. The comparison between kinetic impact and average $\Phi$-values leads on average to a correlation coefficient of 0.62 for all 12 proteins with polarized $\Phi$-value distributions (see Fig.~\ref{correlationCoefficients}), and to an average correlation coefficient of 0.74 if three proteins with negative average $\Phi$-values are excluded. The three proteins have negative average $\Phi$-values below -0.1 in one of the secondary elements, which are difficult to interpret \cite{ozkan01,li00}. The remaining proteins have diffuse $\Phi$-value distributions with similar average $\Phi$-values in the secondary elements. In agreement with the experiments, the distribution of kinetic impact for these proteins is also diffuse. The model thus shows that the polarized of diffuse shapes of most averaged $\Phi$-value distributions can be traced back  to native-state topologies. 

The minimum-ECO-routes defined here represent maybe the simplest possible topology-based modeling of protein folding routes. The prediction of these routes requires the definition of contact maps and contact clusters, but no parameter fitting since the routes are defined as minima of a loop-closure cost function in the space of possible folding sequences. Why can such a simple prediction of folding routes, in combination with a few rules for estimating the kinetic impact of structural elements, reproduce central aspects of mutational experiments? The reason seems to be that the barrier for protein folding is entropic. Furthermore, the relevant entropy here should be loop-closure entropy, since other entropic contributions like the entropy-loss for side-chain `freezing' in contact clusters, once the  loop is closed, should be rather independent of the specific route, i.e.~of the sequences in which the clusters are formed.  Clearly, this simple modeling has its limitations. The model is limited to average $\Phi$-values for secondary elements, since the realistic modeling of detailed $\Phi$-value distributions  requires also energetic characteristics of the specific mutations \cite{merlo}. Another limitation is that the model can not address folding rates. The modeling of folding rates requires an additional estimate for the intrinsic, route-independent formation times of the contact clusters, besides loop-closure. In a previous related model, these intrinsic cluster formation times have been estimated via the number of steps required for `zipping up', or `propagating', a contact cluster after the initial loop-closure step \cite{weikl03a,weikl03b}.  This previous model has five parameters, which were fitted to the folding rates of 24 two-state folders, and considers a more complex set of partially formed zipping states of the clusters.

The model presented here is purely topology-based in the sense that it does not use any sequence-specific information. A central question is whether purely topology-based models can account for the experimentally observed differences in the folding kinetics between proteins with similar folds, or similar overall fold topology. A famous example are protein L and G, which both have a central $\alpha$-helix and two rather symmetric $\beta$-hairpins at the chain ends. Intriguingly, the structural symmetry is `broken' in the $\Phi$-value distributions, and in each of these protein in a different way:  protein L has the largest $\Phi$-values in the N-terminal hairpin \cite{kim00}, and protein G in the C-terminal hairpin \cite{mccallister00}. In the model presented here, native-state topology is captured by the topology of the native contact maps. Protein L and G have very similar folds, but nonetheless small differences in their contact maps. This leads to different folding routes which reproduce the observed symmetry breaking for the two proteins. Other groups have used sequence-specific interaction energies in topology-based models to account for these differences between protein L and G \cite{karanicolas02,clementi03,kameda03,brown04}. The present model traces the symmetry breaking of protein L and G back to native-state topology, but suggests that sequence-specific energetic contributions may affect the folding kinetics of Sso7d and CspB. According to the model, these proteins are governed by parallel folding processes with similar loop-closure cost. However, the experimental $\Phi$-value distributions seem to indicate that one of the parallel processes dominates the kinetics, presumably due to specific energetic interactions which are not considered in the model.

\section{The model}

\subsection{Folding routes}

\begin{figure}
\resizebox{\linewidth}{!}{\includegraphics{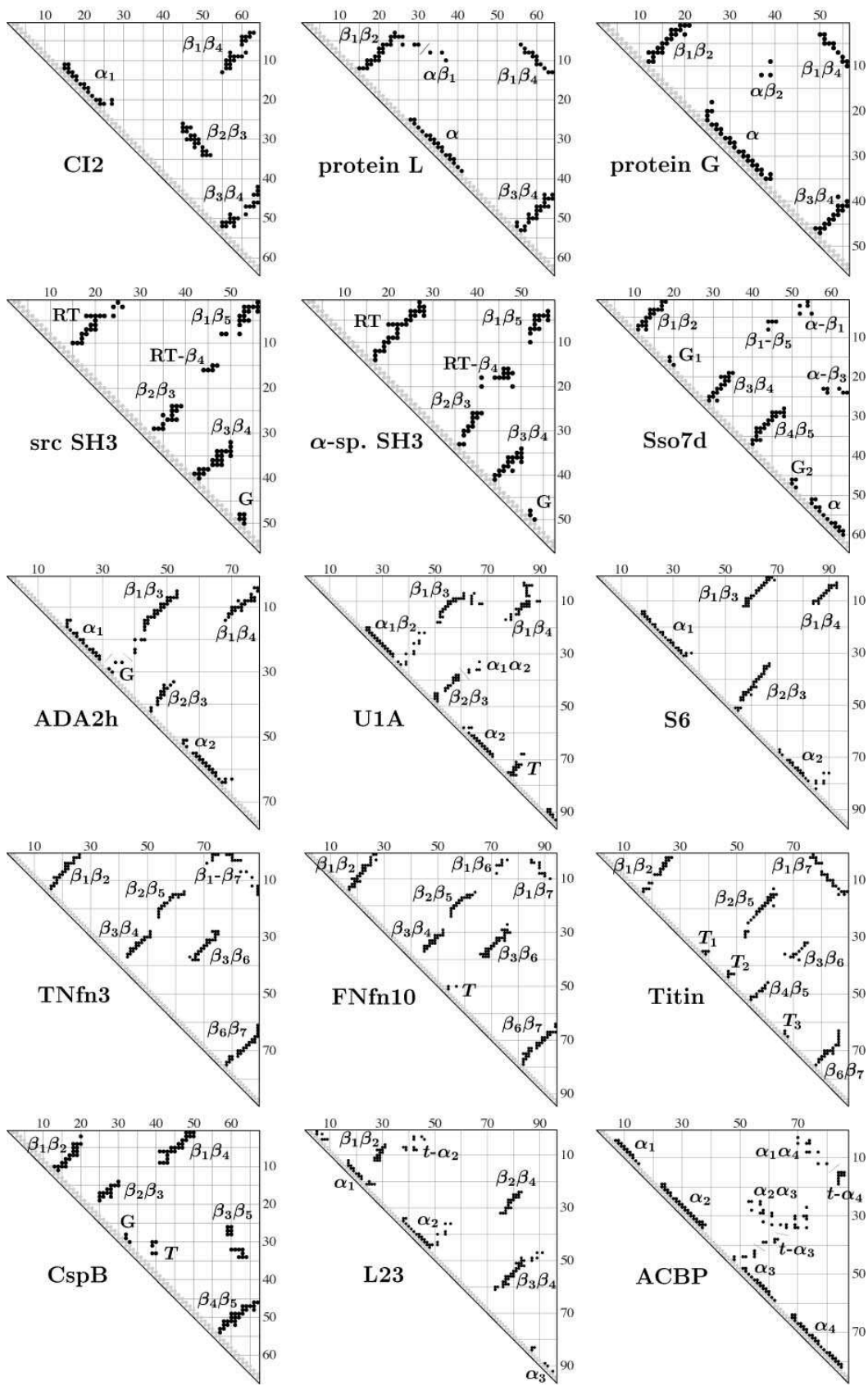}}
\caption{Contact maps and contact clusters of the 15 proteins considered here. \label{contactMaps}}
\end{figure}

The starting point of the model are native contact maps. The native contact map of a protein is a matrix in which the element $(i,j)$ equals 1 if the two residues $i$ and $j$ are in contact in the native structure, and 0 otherwise. Here, two residues are taken to be in contact if the distance between their $C_\alpha$ or $C_\beta$ atoms is less than 6 \AA, and if they are not nearest or next-nearest neighbors in the sequence. The native contacts are grouped into contact clusters (for details, see Methods section). These contact clusters correspond to the structural elements of the protein: helices, $\beta$-strand pairings, and tertiary interactions of helices or $\beta$-sheets. The contact maps and contact clusters of the 15 proteins considered here are shown in Fig.~\ref{contactMaps}. The contact clusters of a protein can be divided into local and nonlocal clusters. Local clusters contain at least one local contact  $(i,j)$ with small contact order CO =$|i-j|<10$, whereas nonlocal clusters do not contain any such local contacts.  

In the model, folding routes are derived from the loop-closure dependencies between the contact clusters. 
To determine the loop-closure relations, all possible sequences are considered in which the clusters can be formed. For a nonlocal cluster, the length of the loop which has to be closed to form the cluster contacts depends on these sequences of cluster formation. In other words, it depends on which other clusters have been formed previously. A simple example with only four contact clusters is CI2 (see Fig.~\ref{contactMaps}). The $\beta_1\beta_4$ cluster of CI2 consists of nonlocal contacts between the two chain ends. Forming these contacts from the fully unfolded state requires the closure of a relatively large loop, and hence costs a large amount of loop-closure entropy. However, forming one or several of the structural  elements $\alpha$, $\beta_2\beta_3$, or $\beta_3\beta_4$ prior to $\beta_1\beta_4$ brings the two chain ends into closer spatial proximity and reduces the length of the loop which has to be closed to form $\beta_1\beta_4$. 

The length of the loop which is closed to form a specific contact between two residues is estimated here via the concept of {\em effective contact order} (ECO) \cite{fiebig93,dill93}. The ECO of the contact is the number of steps along the shortest path between these two residues. Each `step' either is (i) a covalent bond between consecutive  residues in the chain, or (ii) a previously formed noncovalent contact (see Fig.~\ref{eco}). In contrast, the contact order (CO) only takes into account steps of type (i) and hence measures the sequence separation of the two residues. Unlike the ECO, the CO is independent of the folding route, the sequence in which contacts are formed.

The key assumption of the model is that folding routes which involve only closures of relatively small loops dominate the folding process. These routes minimize the entropic loop-closure barriers during folding. To determine the minimum-entropy-loss routes, all possible sequences of cluster formation are considered. Sequences of cluster formation here are called {\em folding sequences}. The formation of each contact cluster in a folding sequence requires to close a loop. The length of this loop is estimated as the minimum ECO among all cluster contacts, the {\em cluster ECO}. The cluster ECO thus is an estimate for the length of the shortest loop that has to be closed to form the cluster in a given partially folded conformation\footnote{More precisely, the cluster ECO is an estimate for the length of the shortest loop that has to be closed to `initiate' the cluster, i.e.~to form the cluster contact(s) with minimum ECO.  After `initiation', the cluster is thought to be `zipped up' in a series of small-loop-closure steps \cite{weikl03a,weikl03b}. These zipping steps do not depend on the folding sequence. Therefore, they are not considered here.}. 
Suppose we have a sequence of clusters $C_1C_2  \ldots  C_n$. Since no contacts have been formed prior to $C_1$,  the ECO $\ell_1$ of this cluster simply is the minimum CO among the cluster contacts. For the other clusters $C_i$ in the folding sequence, the cluster ECO is the minimum ECO among the cluster contacts, given the contacts of the previously formed clusters $C_1, C_2, \ldots, C_{i-1}$. This leads to a sequence of cluster ECOs, or loop lengths, $\ell_1, \ell_2, \ldots, \ell_n$. 

For each folding sequence  $C_1C_2  \ldots  C_n$, the total loop-closure cost can be defined as 
$s=\sum_{i=1}^n  f(\ell_i)$ where $\ell_i$ are the cluster ECOs along the sequence, and $f(\ell_i)$ is a weighting function which increases with the loop length $\ell_i$. For simplicity, the linear weighting function $f(\ell_i)= \ell_i$ is used here \cite{weikl04}. This linear approximation for the free-energy cost of loop closure is not unreasonable since the range of relevant ECOs here only spans roughly one order of magnitude, from 2 to 20 or 30 (see Table 1). The total loop-closure cost then simply is the sum of ECOs  $s=\sum_{i=1}^n  \ell_i$ for all clusters along the sequence.  

The {\em minimum-ECO sequences} to a given cluster $C_n$ are simply defined as local minima of the loop-closure cost $s$ in the space of all possible folding sequences to $C_n$. In this space, the neighbors of a given folding sequence $C_1C_2  \ldots  C_n$ are those sequences which are obtained either by deleting one or several of the clusters from  $C_1C_2  \ldots  C_n$, or by adding one or several `new' clusters somewhere in the sequence (see also Methods section). In principle, two neighboring folding sequences can have the same local minimum value of $s$. In this case, the longer sequence among the two is selected as the minimum-ECO sequence.  

Finally, all minimum-ECO sequences which consist of the same set of clusters are taken to represent the same {\em minimum-ECO route}. These sequences have the same loop-closure cost $s$ and differ only by permutations from each other, which indicates parallel folding processes on the route. Suppose the ECO of the nonlocal cluster $C_3$ is only affected by the two local clusters $C_1$ and $C_2$. Since the ECOs of the local clusters $C_1$ and $C_2$ are independent of each other, the two sequences $C_1C_2C_3$ and $C_2C_1C_3$ then both are minimum-ECO sequences, representing the same minimum-ECO route. On this minimum-ECO route, the two local clusters $C_1$ and $C_2$ form in parallel, prior to the nonlocal cluster $C_3$.

Table 1 summarizes the loop-closure hierarchies on the minimum-ECO routes for the proteins considered here. For each nonlocal cluster of a protein, all clusters formed prior on the minimum-ECO route are shown. For some nonlocal clusters, there are multiple minimum-ECO routes. These multiple routes correspond to different local minima of the loop-closure cost $s$ in the space of folding sequences. However, local minima with a loop-closure cost $s$ which is by 10 or more larger than the global minimum are neglected. These local minima represent folding routes with significantly larger entropic barriers.

\subsection{Kinetic impact of secondary structural elements}

The most important kinetic data for two-state folders are $\Phi$-values, which reflect the impact of mutations on the folding kinetics (see eq.~(\ref{phi})).  A $\Phi$-value distribution for a protein is obtained by considering many single-residue mutations throughout the protein chain. For comparison with the model, the experimental $\Phi$-value distributions here are averaged over whole secondary structural elements (helices or sheets). These average $\Phi$-values typically are positive and indicate the relative `kinetic importance', or `kinetic impact', of the secondary structural elements. For example, a relatively large average $\Phi$-value for a secondary element indicates that mutations in this element have a strong impact on the folding rate.

In order to compare with average experimental $\Phi$-values, the kinetic impact of secondary structural elements here is estimated from the loop-closure hierarchies summarized in Table 1. For this purpose, we first have to the consider the kinetic impact of the contact clusters. In a semi-quantitative approach, the kinetic impact of contact clusters and secondary structural elements here is divided into high (H), medium (M), or low (L). 

First, it seems reasonable to assume that the kinetic impact of a cluster should be related to how often it appears on the minimum-ECO routes to other clusters. Suppose a local cluster appears on minimum-ECO routes to all non-local clusters. Mutations affecting the formation of this cluster then should strongly affect the overall folding kinetics. Hence, the cluster has a high kinetic impact. To quantify this notion, the {\em occurrence number} $n$ of a cluster is defined as the number of times it appears on all routes to all (other) nonlocal clusters. In other words, $n$ simply is the number of times the cluster occurs in the third column of Table 1. In terms of occurrence numbers, the first rule is:
\begin{enumerate}
\item[(1)] The kinetic impact of a cluster is high (H) if its occurrence number $n$ on the minimum-ECO routes is larger than or equal to $\frac{2}{3} n_\text{max}$. Here, $n_{max}$ is the maximum value of $n$ among all clusters of the protein. The impact of the cluster is medium (M) for $\frac{1}{3}n_\text{max}\le n <\frac{2}{3}n_\text{max}$. The impact is low (L) for $n <\frac{1}{3}n_\text{max}$.
\end{enumerate}

Second, the kinetic impact of nonlocal clusters should also be affected by the cluster ECO. Suppose a nonlocal cluster has a high cluster ECO on all minimum-ECO routes. This means that forming the cluster always involves the closure of a relatively large loop. It seems reasonable to assume that the kinetic impact of the cluster then is high, since the contacts of these clusters have to balance a relatively high loop-closure entropy. In other words, the formation of the cluster and, hence, the overall folding kinetics should be highly sensitive to mutations affecting the cluster contacts. The second rule is: 
\begin{enumerate}
\item[(2)] A nonlocal cluster has a high (H) kinetic impact if the ECO of this cluster is larger than 10 an all routes. The kinetic impact is medium (M) if the smallest cluster ECO  has a value from 6 to 10, unless rule (1) specifies high impact. 
\end{enumerate}
According to the rules (1) and (2), the kinetic impact of a cluster thus is low if its occurrence number is small, and the cluster ECO is not larger than 5.

Finally, suppose a protein has two nonlocal clusters $C_1$ and $C_2$ which fold in parallel. This means that the cluster $C_1$ does not appear on the minimum-ECO routes to $C_2$, and vice versa. In general, the loop-closure cost for forming, e.g, $C_1$ can be significantly larger than the loop-closure cost for forming $C_2$.  It seems reasonable that clusters appearing on the minimum-ECO routes to $C_1$ should then have a higher kinetic impact than clusters appearing only on minimum-ECO routes to $C_2$, since  the entropic loop-closure barrier for forming $C_1$ is significantly larger. Therefore, the third rule is:
\begin{enumerate}
\item[(3)] If two nonlocal clusters $C_1$ and $C_2$  do not occur on minimum-ECO routes to other clusters and have minimum loop-closure costs $s_1$ and $s_2$ with $s_1 > s_2 + 5$, the cluster occurrences on the routes to $C_2$ are not taken into account in rule (1). In particular, clusters which appear only on routes to $C_2$ have a low kinetic impact, independent of their ECO. 
\end{enumerate}

The rules (1), (2), and (3) define the kinetic impact of clusters. The translation into kinetic impact of secondary elements (strands or helices) is straightforward. The kinetic impact of a secondary element is high (H)  if it  has contacts in a cluster with high kinetic impact, and low (L) if it only has contacts in clusters with low kinetic impact. The kinetic impact of a secondary element is medium (M) if it has contacts in clusters with medium kinetic impact, but no contacts in clusters with high kinetic impact. As an example, the high kinetic impact of the clusters $\alpha_i$ and $\beta_k\beta_l$ of a protein results in a high kinetic impact of the secondary elements $\alpha_i$, $\beta_k$, and $\beta_l$. The relation between secondary elements and contact clusters is summarized in the cluster labels of Fig.~\ref{contactMaps}. 
  
Table 2 shows average experimental $\Phi$-values  and kinetic impact for the strands and helices of the 15 proteins considered here. To illustrate the rules (1) and (2), consider for example the src SH3 domain. This protein has two nonlocal clusters, RT-$\beta_4$ and $\beta_1\beta_5$ (see Fig.~\ref{contactMaps}). The clusters $\beta_2\beta_3$ and $\beta_3\beta_4$ appear on the minimum-ECO routes  to both nonlocal clusters (see Table 1) and, hence, have the occurrence number 2. The cluster RT only appears on the route to $\beta_1\beta_5$ and, hence, has occurrence number 1. According to rule (1), the kinetic impact of $\beta_2\beta_3$ and $\beta_3\beta_4$ thus is high (H), and the kinetic impact of RT is medium (M). According to rule (2), the kinetic impact of the cluster RT-$\beta_4$ is medium since it has the cluster ECO 10. Finally, the kinetic impact of $\beta_1\beta_5$ is low (L) since it has a small cluster ECO of 5 and occurrence number 0. Therefore, the kinetic impact of the strands $\beta_2$, $\beta_3$, and $\beta_4$ is high, the kinetic impact of RT is medium, and the kinetic impact of $\beta_1$ and $\beta_5$ is low, in perfect agreement with the average $\Phi$-values (see Table 2).

Rule (3) affects the proteins U1A and L23. In the case of U1A, the cluster $\alpha_1\alpha_2$ does not occur on the minimum-ECO routes to the two other nonlocal clusters $\beta_1\beta_3$ and $\beta_1\beta_4$ and has a significantly smaller loop-closure cost than  $\beta_1\beta_4$. Therefore, $\alpha_2$ has a low kinetic impact, since it only appears on the minimum-ECO route to $\alpha_1\alpha_2$. In the case of L23, the nonlocal clusters t-$\alpha_2$ folds in parallel to $\beta_2\beta_4$, with significantly smaller loop-closure cost.  As a consequence, the kinetic impact of $\beta_1$ and $\alpha_1$ is low since these secondary elements are only involved in the folding of  t-$\alpha_2$.

\section{Results and discussion}

To evaluate the model, is is useful to distinguish between proteins with polarized and diffuse $\Phi$-value distributions. Here, this distinction is based on the average $\Phi$-values for the helices and strands.  A distribution is polarized if the $\Phi$-values for some of the secondary elements are significantly larger than for other secondary elements. To quantify this notion, a $\Phi$-value distribution here is defined as polarized if at least two average $\Phi$-values are by more than a factor 2.5 smaller than the maximum value of the distribution. A  $\Phi$-value distribution is diffuse if this is not the case. In a diffuse distribution, all or all except one of the average $\Phi$-values are larger than 40\% of the maximum among these values. An analogous definition can also be applied to the distribution of kinetic impact derived from the minimum-ECO routes. The distribution is diffuse if all or all except one of the secondary elements have high kinetic impact.

\begin{figure}
\begin{center}
\resizebox{0.8\linewidth}{!}{\includegraphics{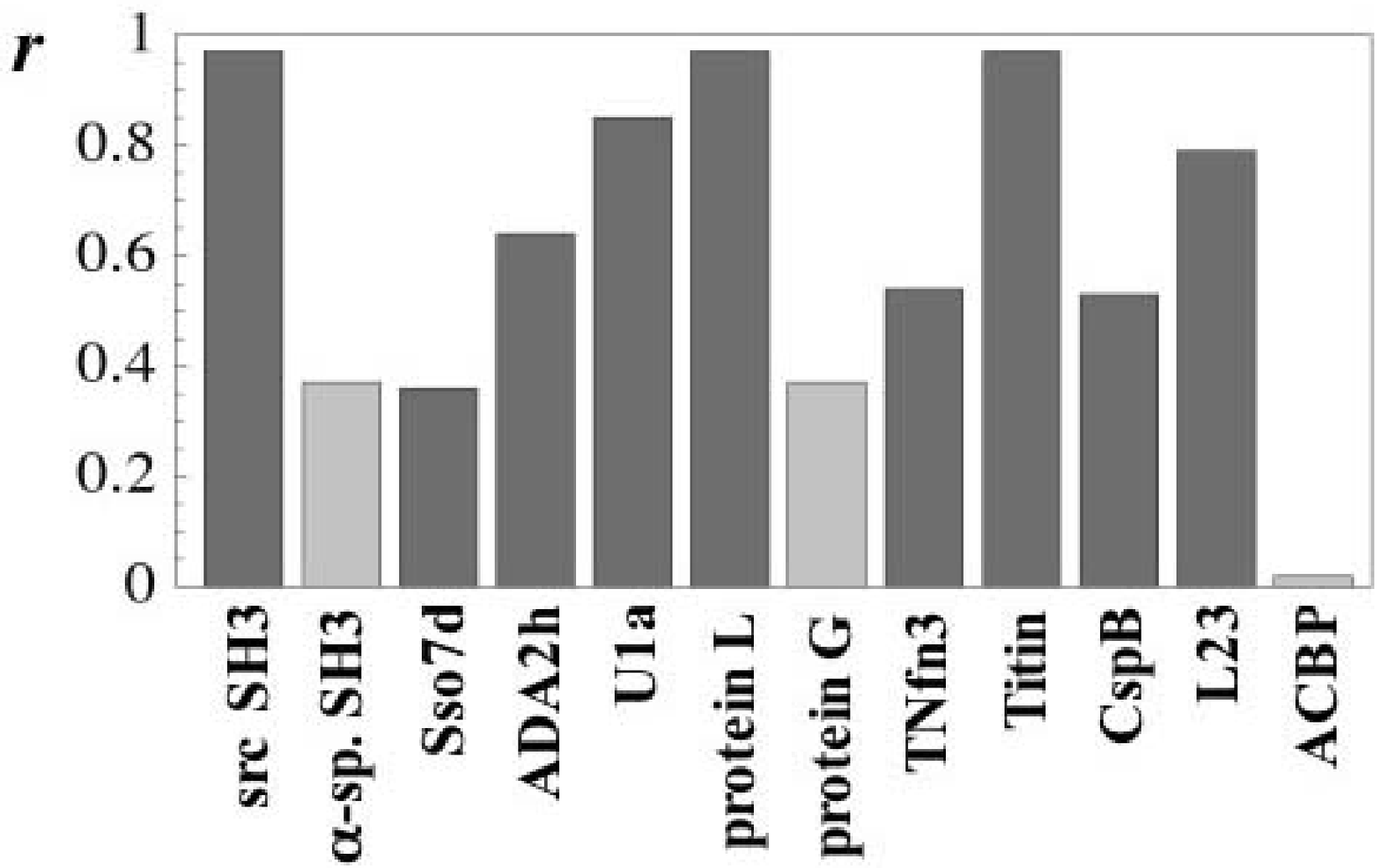}}
\end{center}
\vspace{0.5cm}
\caption{Correlation cofficients $r$ for the comparison between average experimental $\Phi$-values and kinetic impact of the 12 proteins with polarized $\Phi$-value distributions. The light grey bars represent the correlation coefficients of proteins with negative average $\Phi$-values below -0.1 in one of the secondary elements. On average, the correlation coefficient is 0.62 for all 12 proteins, and 0.74 for the 9 proteins with positive $\Phi$-values. For U1A, the correlation coefficients for the two $\Phi$-value distributions at $\beta=0.5$ and $\beta=0.7$ (see Table 2) are 0.91 and 0.79. Here, the average 0.85 of these two values is presented. -- To test the statistical significance of the observed correlations, one can compare the obtained correlation coefficient with those between the theoretical distribution and all possible random permutations of the experimental distribution for each of the proteins. The fraction $p$ of random permutations of the experimental data which have an equally high or larger correlation coefficient with the theoretical distribution can be interpreted as probability to obtain the correlations shown in the Figure, or larger ones, by chance.
This probability is $p=0.017$ for src SH3, $p=0.20$ for $\alpha$-spectrin SH3, $p=0.10$ for Sso7d, $p=0.17$ for ADA2h, $p=0.033$ and 0.067 for U1A, $p= 0.033$ for protein L, $p=0.30$ for protein G, $p=0.29$ for TNfn3, $p=0.026$ for Titin, $p=0.17$ for CspB, and $p=0.036$ for L23. Despite the relatively small number of data points (the proteins have between 4 and 7 secondary structural elements), the obtained correlations are statistically significant. The probability $p$ for obtaining an average correlation coefficient of 0.62 or larger for all 12 proteins by chance is smaller than $10^{-6}$.
\label{correlationCoefficients}}
\end{figure}

According to this definition, 3 among the 15 proteins considered here have a diffuse $\Phi$-value distribution. These proteins are CI2, S6, and FNfn10. In agreement with the experiments, the distribution of kinetic impact for the secondary structural elements of these proteins is also diffuse (see Table 2). The remaining 12 proteins have polarized $\Phi$-value distributions. Fig.~\ref{correlationCoefficients} shows the correlation coefficient $r$ between average ${\Phi}$-values and kinetic impact for each of these proteins. The calculate the correlation coefficients, the values 0, 1, and 2 are assigned to the kinetic impact L, M, and H.\footnote{Any other `equidistant' values $a$, $a+b$, and $a+2 b$ with $b>0$ for the kinetic impact L, M, and H result in the same correlation coefficients. Correlation coefficients are only given for proteins with polarized distributions since the do not  reflect the quality of the modeling in the case of diffuse distributions with rather similar average $\Phi$-values for the secondary elements.} The correlation coefficient $r$ can attain values in the range -1 to 1 where 1 means `perfect' correlation (proportionality), 0 means no correlation, and negative values mean anticorrelation. 
 
Three of the lowest correlation coefficients are obtained for the $\alpha$-spectrin SH3 domain, protein G, and ACBP (see Fig.~\ref{correlationCoefficients}). These proteins have clearly negative average $\Phi$-values (smaller than -0.1) in one of the secondary elements. For the comparison with kinetic impact, the negative average $\Phi$-values were simply taken to be zero. However, excluding the helix $\alpha_2$ of ACBP with negative average $\Phi$-value from the correlation analysis leads to a correlation coefficient of 0.94, instead of 0.02. For the $\alpha$-spectrin SH3 domain, excluding the strand $\beta_2$ from the comparison leads to a correlation coefficient of 0.69 instead of 0.37. Thus,  the relatively low correlation coefficients for these two proteins can be traced back directly to the secondary elements with negative $\Phi$-values. 
  
Two other proteins with relatively low correlation coefficients in Fig.~\ref{correlationCoefficients} are Sso7d and CspB. These proteins have in common that the nonlocal clusters fold in parallel on the minimum-ECO routes.  In the case of CspB, the nonlocal clusters are $\beta_1\beta_4$ and $\beta_3\beta_5$. Since the total loop-closure cost of the two parallel folding processes leading to these clusters are similar (see Table 1), the model takes them to be equally important for the kinetics. However, the experimental $\Phi$-values seem to indicate that the folding process leading the  $\beta_1\beta_4$ has a larger impact on the kinetics than the parallel process leading to  $\beta_3\beta_5$. The strands $\beta_1$ to $\beta_3$ of the two clusters $\beta_1\beta_2$ and $\beta_2\beta_3$ which are formed prior to $\beta_1\beta_4$  have relatively large average $\Phi$-values. In contrast, the strands  of the cluster $\beta_4\beta_5$, which is formed prior to $\beta_3\beta_5$ on the parallel folding process, have significantly smaller average $\Phi$-values. In the case of Sso7d, the three nonlocal clusters $\alpha$-$\beta_3$, $\beta_1$-$\beta_5$ and $\alpha$-$\beta_1$ fold in parallel, with comparable loop-closure cost. Here, the experimental $\Phi$-values seem to indicate that the folding process leading to $\alpha$-$\beta_3$ dominates the folding kinetics. According to the model, the clusters formed prior to $\alpha$-$\beta_3$ are $\beta_3\beta_4$, $\beta_3\beta_5$, and $\alpha$. The secondary elements of these clusters have medium are large average $\Phi$-values, whereas the $\Phi$-values of the remaining secondary structural elements $\beta_1$, $\beta_2$, and $G_1$ are significantly smaller. In both proteins, specific energetic interactions, which are not taken into account in the model, may be responsible for the dominance of one the parallel folding processes with similar entropic loop-closure barriers. 
 
For the remaining majority of proteins, the model reproduces the polarized $\Phi$-value distributions with relatively large correlation coefficients. This shows that $\Phi$-value distributions averaged over secondary elements are dominated by native-state topology. In the model, the native-state topology is captured by the topology of the native contact maps, or more precisely, by the ECO-dependencies between the contact clusters. Interestingly, the model is able to reproduce the experimentally observed differences in the $\Phi$-value distributions of protein L and G without sequence-specific information.  These two proteins have very similar folds, but nonetheless small differences in their contacts maps. Whereas protein L has a small tertiary $\alpha\beta_1$ cluster, protein G has a tertiary $\alpha\beta_2$ cluster. This results in different folding routes and different distributions of kinetic impact (see Tables 1 and 2). In the case of protein L, the N-terminal hairpin $\beta_1\beta_2$  has higher kinetic impact the C-terminal hairpin $\beta_3\beta_4$, in agreement with the average $\Phi$-values. In the case of protein G, the kinetic impact and average $\Phi$-values are larger for the C-terminal hairpin $\beta_3\beta_4$. Other groups have used sequence-specific interaction energies to reproduce these differences between protein L and G \cite{karanicolas02,clementi03,kameda03,brown04}. 

The folding routes of the model are hierarchic in the sense that the formation of nonlocal structural elements typically requires the prior formation of other, more local structural elements. It is important to note that the hierarchic folding routes do not contradict cooperative two-state folding with a characteristic single-exponential relaxation dynamics. We have recently developed a free-energy based model with similar loop-closure dependencies \cite{weikl04}. In this model, two-state folding cooperativity is reproduced when assuming that the local structural elements are unstable. On one hand, the nonlocal structural elements then stabilize the overall fold and, thus, also  the local elements. On the other hand, the local structural elements reduce the  loop-closure entropies for forming the nonlocal elements. \footnote{Similar in spirit, the diffusion-collision model of Karplus and Weaver assumes that individual microdomains such as helices are unstable \cite{karplus76,karplus94}. A direct, energetic local-nonlocal coupling has been recently used by Kaya and Chan \cite{kaya03} to obtain two-state cooperitivity in a simple lattice model.} On the energy landscapes, the formation of local structural elements then corresponds to uphill steps in free energy, and the formation of nonlocal structural elements to steps downhill in free energy, with characteristic barrier or `transition' states in between. For an $\alpha$-helical protein, the hierarchy of local and nonlocal structural elements corresponds to a hierarchy of secondary and tertiary elements \cite{baldwin99}, since the local structural elements are individual helices. However, this correspondence is not general: a $\beta$-hairpin, for example, is a local structural element, but involves both secondary and tertiary structure formation.  

\section{Conclusions}

The model presented here derives folding routes of proteins and the `kinetic impact' of secondary structural elements from native structures. In a first step, minimum-entropy-loss routes are derived from the native contact maps. This step reveals characteristic loop-closure dependencies between local and nonlocal structural elements. In a second step, the model estimates the kinetic impact of secondary elements from the folding routes. In a systematic comparison for a large set of small and well-characterized proteins, relatively high correlation coefficients are obtained between kinetic impact and average experimental $\Phi$-values of the secondary elements.  The model thus indicates that the shape of $\Phi$-value distributions is dominated by native-state topology.

\section{Acknowledgements}

I would like to thank Ken Dill for enjoyable and highly stimulating interactions and collaborations.

\section{Methods}

\subsection*{Contact clusters}

The native contacts are grouped into contact clusters. In general, two contacts $(i,j)$ and $(k,l)$ are taken to be in the same cluster if they are close together on the contact map, according to the distance criterion $|i-k| + |j-l|\le 4$. However, peripheral contacts $(i,j)$ which have a minimum distance $|i-k| + |j-l|= 4$ to the other contacts in the cluster are discarded. For clusters corresponding to helices or $\beta$-strand pairings, also contacts $(i,j)$ which have a distance $|i-k| + |j-l|= 3$ to only one contact $(k,l)$ in the cluster and larger distances to the other contacts are defined as peripheral and discarded. The definition of peripheral contacts is more restrictive for these clusters since they are typically more compact than clusters corresponding to tertiary interactions of helices or sheets. A cluster has to contain at least three contacts. Isolated contacts or contact pairs are not taken into account. Discarding peripheral and isolated contacts helps to avoid an unreasonably large impact of individual contacts on the cluster ECOs, and hence on the model results. 

The following PDB files have been used to determine the contact maps and contact clusters: CI2 (1COA); protein L (2PTL, residues 15 to 78); protein G (1PGB), src SH3 domain (1SRL); $\alpha$-spectrin SH3 domain (1SHG); ADA2h (1AYE); U1A (1URN, chain A); S6 (1RIS); TNfn3 (1TEN); FNfn10 (1FNF, residues 1416 to 1509); Titin (1TIT); CspB (1CSP); L23 (1N88); ACBP (2ABD). 

\subsection*{Minimum-ECO sequences}

As defined in section 2, the minimum-ECO sequences to a given cluster $C_n$ locally minimize the loop-closure cost $s$ in the space of folding sequences. Starting with the set of all possible folding sequences to $C_n$, the minimum-ECO sequences are obtained by applying the following two rules.
\begin{enumerate}

\item[(1)] If two folding sequences $a$ and $b$ have the loop-closure costs $s_a<s_b$ and the set of clusters $\{C_1^{(a)},C_2^{(a)},\ldots C_{m}^{(a)}, C_n\}$ of sequence $a$ is a subset of the clusters  $\{C_1^{(b)},C_2^{(b)},\ldots C_{k}^{(b)}, C_n\}$ of sequence $b$, then sequence $b$ is discarded.

\item[(2)] If two folding sequences $a$ and $b$ have the loop-closure costs $s_a\ge s_b$ and the set of clusters $\{C_1^{(a)},C_2^{(a)},\ldots C_{m}^{(a)}, C_n\}$ of route $a$ is a proper subset of the clusters  $\{C_1^{(b)},C_2^{(b)},\ldots C_{k}^{(b)}, C_n\}$ of sequence $b$, then sequence $a$ is discarded. 
\end{enumerate}

The rules (1) and (2) are best illustrated in a simple example. Suppose the folding sequence $C_1C_2C_3$  to cluster $C_3$ has the loop-closure cost $s_a$.  Suppose now that the cluster $C_0$ is a cluster which does not affect any of the cluster ECOs of  $C_1$, $C_2$, or $C_3$. If the cluster $C_0$ is, e.g., a local cluster with small cluster ECO the cost $s_b$ of the sequence $C_0C_1C_2C_3$ is only slightly larger  than the cost $s_a$ of the sequence $C_1C_2C_3$. However, since there is no ECO-dependence between $C_0$ and the other three clusters, the sequence $C_0C_1C_2C_3$ is not a reasonable candidate for a minimum-ECO sequence to cluster $C_3$, and hence is discarded by rule (1). 

On the other hand, let's suppose that the sequence $C_1C_3$ has a larger cost $s_b$ than the sequence $C_1C_2C_3$. This means that the prior formation of $C_2$ affects the ECO of $C_3$. Therefore, the sequence $C_1C_3$ is discarded by rule (2) from the possible minimum-ECO sequences to $C_3$.

\subsection*{Secondary structure classification}

The calculation of average experimental $\Phi$-values for helices and strands requires secondary structure classifications. Where possible, the secondary structure definitions given in the PDB files (see above) have been used here. The PDB files of TNfn3 and the $\alpha$-spectrin SH3 domain do not contain secondary structure classifications. For TNfn3 and the structurally analogous protein FNfn10, secondary structure classifications have been taken from Hamill et al\cite{hamill00} and Cota et al\cite{cota01}. For the $\alpha$-spectrin SH3 domain, the secondary structure definition of the DSSP algorithm \cite{kabsch83} has been used. In the case of CspB, the first two substrands given in the PDB file are combined into strand $\beta_1$,  the last two substrands into $\beta_5$, and the $3_{10}$ helix is defined from residues 30 to 33. The RT loop of the two SH3 domains is an irregular secondary structure defined here from residues 14 to 26.

\newpage

\begin{table}

\vspace*{-0.9cm}
Table 1: Loop-closure events on minimum-ECO routes \\[0.1cm]

{\renewcommand{\baselinestretch}{0.65}\footnotesize
\begin{tabular}{cclcc}
             & nonlocal &                                     &  \hspace*{0cm}ECO for non-           &  loop-closure \\[-0.05cm]
protein &  cluster   & clusters formed prior &  \hspace*{0cm}local cluster & cost \\
\hline\hline
CI2 &  $\beta_2\beta_3$ &    ---                                                                               & 16 & 16 \\
               & $\beta_1\beta_4$ &$\alpha_1$, $\beta_3\beta_4$, $\beta_2\beta_3$,  & 7 & 27    \\
\hline 
protein L & $\alpha\beta_1$ &$\beta_1\beta_2$                                     & 6 & 9 \\
                 & $\beta_1\beta_4$ &$\beta_1\beta_2$, $\alpha$, $\beta_3\beta_4$, $\alpha\beta_1$ & 9 & 22 \\ 
\hline                   
protein G & $\alpha\beta_2$ &$\alpha$                                                  & 10 & 13 \\
                 & $\beta_1\beta_4$ &$\beta_1\beta_2$, $\alpha$, $\beta_3\beta_4$& 9 & 18 \\
                 &                                   &or: $\alpha$, $\beta_3\beta_4$, $\alpha\beta_2$ & 3 & 19\\
\hline

src SH3 &  RT-$\beta_4$ &$\beta_2\beta_3$, $\beta_3\beta_4$        & 10 & 17 \\
                       &  $\beta_1\beta_5$ &RT, $\beta_2\beta_3$, $\beta_3\beta_4$ & 5 & 17\\
\hline               
$\alpha$-spSH3 &  RT-$\beta_4$ &$\beta_2\beta_3$    & 7 & 10 \\
                       &  $\beta_1\beta_5$ &RT, $\beta_2\beta_3$, $\beta_3\beta_4$, G & 5 & 17\\
                       &                                 &or: RT, $\beta_2\beta_3$, G, RT-$\beta_4$ & 3 & 17\\
\hline  
Sso7d   & $\alpha$-$\beta_3$ &$\beta_3\beta_4$, $\beta_4\beta_5$, $\alpha$ & 7 & 16 \\
                      & $\beta_1$-$\beta_5$ &$\beta_1\beta_2$, $\beta_3\beta_4$, $\beta_4\beta_5$ & 9 & 18 \\
                      & $\alpha$-$\beta_1$ &$\beta_1\beta_2$, $\beta_3\beta_4$, $\beta_4\beta_5$ & 12 & 21 \\
\hline                      
ADA2h & $\beta_1\beta_3$  & G                                                     & 8 & 11 \\
                     &  $\beta_1\beta_4$ &$\alpha_1$, G, $\beta_2\beta_3$, $\alpha_2$ & 9 & 21 \\
                     &                                  &or: G, $\alpha_2$, $\beta_1\beta_3$ & 7 & 21 \\
\hline     
U1A & $\alpha_1\alpha_2$ &$\beta_2\beta_3$, $\alpha_2$               & 3 & 9 \\
         & $\beta_1\beta_3$   &$\alpha_1\beta_2$, $\beta_2\beta_3$    & 6 & 12 \\
         & $\beta_1\beta_4$  &$\alpha_1\beta_2$, $\beta_2\beta_3$, $T$, $\beta_1\beta_3$ & 2 &  17 \\
\hline
S6          & $\beta_1\beta_3$ & $\alpha_1$, $\beta_2\beta_3$   & 14 & 20 \\
               & $\beta_1\beta_4$ & $\alpha_1$, $\beta_2\beta_3$, $\alpha_2$  & 14 & 23 \\       
\hline
TNfn3      & $\beta_2\beta_5$ & $\beta_3\beta_4$                                    & 9 & 15 \\
                 & $\beta_1$-$\beta_7$ &$\beta_3\beta_4$, $\beta_6\beta_7$, $\beta_2\beta_5$ & 2 & 20 \\
                 & $\beta_3\beta_6$  &$\beta_1\beta_2$,  $\beta_3\beta_4$, $\beta_6\beta_7$, $\beta_2\beta_5$,  $\beta_1$-$\beta_7$ & 4 & 27 \\[0.1cm]
\hline              
FNfn10   & $\beta_2\beta_5$ &$\beta_3\beta_4$                                    & 9 & 17 \\
                & $\beta_3\beta_6$  &$T$                                                                 & 22 & 25 \\
                & $\beta_1\beta_6$ &$\beta_1\beta_2$, $T$, $\beta_3\beta_6$ & 2 & 30 \\
                &                         &or: $\beta_3\beta_4$, $\beta_6\beta_7$, $\beta_2\beta_5$, $\beta_1\beta_7$ & 2 & 31 \\
                & $\beta_1\beta_7$ &$\beta_3\beta_4$, $\beta_6\beta_7$, $\beta_2\beta_5$                                 & 9 & 29 \\
                &                   &or: $\beta_1\beta_2$, $T$, $\beta_6\beta_7$, $\beta_3\beta_6$, $\beta_1\beta_6$   & 2 & 34 \\
\hline
Titin        & $\beta_2\beta_5$ & $T_1$, $T_2$                  & 14 & 20 \\
                & $\beta_3\beta_6$ & $\beta_4\beta_5$, $T_3$ & 14 & 20 \\
                & $\beta_1\beta_7$ & $T_1$, $T_2$, $\beta_6\beta_7$, $\beta_2\beta_5$ & 2 & 25 \\
                &                                &or: $\beta_1\beta_2$, $\beta_4\beta_5$, $T_3$, $\beta_3\beta_6$ & 8 & 31 \\
\hline   
CspB     & $\beta_1\beta_4$ & $\beta_1\beta_2$, $\beta_2\beta_3$, $T$   & 4 & 19 \\
               & $\beta_3\beta_5$ & $\beta_4\beta_5$            & 14 & 17 \\                
\hline               
L23        & $\beta_1\beta_2$ & $\alpha_1$                       & 3 & 6 \\
               & $\beta_3\beta_4$ &  --                   & 12 & 12 \\
               & $t$-$\alpha_2$ & $\alpha_1$, $\alpha_2$, $\beta_1\beta_2$ & 5 & 14 \\
               & $\beta_2\beta_4$ & $\alpha_2$, $\beta_3\beta_4$  & 6 & 21 \\
\hline               
ACBP   & $t$-$\alpha_3$ & $\alpha_3$                     & 5 & 8 \\
              & $\alpha_2\alpha_3$ & $\alpha_2$, $\alpha_3$, $t$-$\alpha_3$ & 3 & 14 \\
              & $t$-$\alpha_4$  & $\alpha_2$, $\alpha_3$, $\alpha_4$,  $t$-$\alpha_3$, $\alpha_2\alpha_3$ & 6 & 23 \\
              & $\alpha_1\alpha_4$ & $\alpha_1$, $\alpha_2$, $\alpha_3$, $t$-$\alpha_3$, $\alpha_2\alpha_3$ & 9 & 26 \\
               &  &or: $\alpha_2$, $\alpha_3$, $\alpha_4$,  $t$-$\alpha_3$, $\alpha_2\alpha_3$, $t$-$\alpha_4$ & 5 & 28 

\end{tabular}}
\vspace{0.5cm} 
\end{table}

\begin{table}
\vspace*{-1cm}
Table 2: Average $\Phi$-values and kinetic impact of secondary structural elements\\[0.3cm]
{
\renewcommand{\baselinestretch}{0.75}\scriptsize
\begin{tabular}{lccccccc}
CI2& $\beta_1$  & $\alpha$  & $\beta_2$ & $\beta_3$ & $\beta_4$ &&\\
$\bar{\Phi}_{\text{exp}}$ & 0.23 (1) & 0.32 (12) & 0.15 (4) & 0.32 (4) & 0.03 (2)&& \\
kinimpact& M & H & H & H & H &&\\[0.3cm]
protein L & $\beta_1$  & $\beta_2$  & $\alpha$ & $\beta_3$ & $\beta_4$ &&\\
$\bar{\Phi}_{\text{exp}}$ & 0.36 (9) & 0.46 (7) & 0.15 (16) & 0.18 (4) & 0.14 (7)&& \\
kinimpact& H & H & M & M & M &&\\[0.3cm]
protein G & $\beta_1$  & $\beta_2$  & $\alpha$ & $\beta_3$ & $\beta_4$ &&\\
$\bar{\Phi}_{\text{exp}}$ & 0.36 (3) & $-0.16$ (4) & 0.13 (9) & 0.63 (2) & 0.27 (4)&& \\
kinimpact& M & M & H & H & H &&\\[0.3cm]
src SH3 & $\beta_1$  & RT & $\beta_2$ & $\beta_3$ & $\beta_4$ & G & $\beta_5$ \\
$\bar{\Phi}_{\text{exp}}$ & 0.02 (4) & 0.10 (8) & 0.46 (3) & 0.53 (6) & 0.43 (6)& -- & $-0.04$ (2) \\
kinimpact& L & M & H & H & H & L & L \\[0.3cm]
$\alpha$-spec SH3 & $\beta_1$  & RT & $\beta_2$ & $\beta_3$ & $\beta_4$ & G & $\beta_5$  \\
$\bar{\Phi}_{\text{exp}}$ & 0.08 (2) & 0.26 (3) & $-0.20$ (1) & 0.66 (3) & 0.60 (2)& 0.53 (1) & 0.16 (1) \\
kinimpact & L & H & H & H & M & H & L  \\[0.3cm]
Sso7d & $\beta_1$  & $\beta_2$ & $G_1$ & $\beta_3$ & $\beta_4$ & $\beta_5$ & $\alpha$ \\
$\bar{\Phi}_{\text{exp}}$ & $-0.03$ (2) & 0.11 (2) & $-0.03$ (2)& 0.96 (2) & 0.27 (4) & 0.19 (5)& $0.41$ (4)  \\
kinimpact & H & H & L & H & H & H & H \\[0.3cm]
ADA2h & $\beta_1$  & $\alpha_1$ & G & $\beta_2$ & $\beta_3$ & $\alpha_2$ & $\beta_4$  \\
$\bar{\Phi}_{\text{exp}}$ & $0.42$ (3) & 0.26 (3) & -- & 0.06 (2) & 0.29 (3) & 0.49 (4)& $0.14$ (2) \\
kinimpact & M & M & H & M & M & H & M \\[0.3cm]
U1A & $\beta_1$  & $\alpha_1$ & $\beta_2$ & $\beta_3$ & $\alpha_2$ & $\beta_4$ & \\
$\bar{\Phi}_{\text{exp}}$ ($\beta=0.5$) & $0.23$ (2) & 0.38 (3) & 0.73 (3) & --  & 0.00 (1)& $0.00$ (1)& \\
$\bar{\Phi}_{\text{exp}}$ ($\beta=0.7$) & $0.43$ (2) & 0.63 (3) & 0.98 (3) & --  & 0.50 (1)& $0.23 $ (1)& \\
kinimpact & M & H & H & H & L & L &\\[0.3cm]
S6 & $\beta_1$  & $\alpha_1$ & $\beta_2$  & $\beta_3$ & $\alpha_2$ & $\beta_4$ &\\
$\bar{\Phi}_{\text{exp}}$ & 0.34 (4) & 0.25 (4) & 0.24 (1) & 0.31 (5) & 0.28 (2)&0.14 (2) & \\
kinimpact& H & H & H & H & M & H &\\[0.3cm]
TNfn3 & $\beta_1$  & $\beta_2$  & $\beta_3$ & $\beta_4$ & $\beta_5$ & $\beta_6$ & $\beta_7$\\
$\bar{\Phi}_{\text{exp}}$ & 0.12 (3) & 0.27 (2) & 0.36 (3) & 0.55 (2) & 0.47 (2)& 0.42 (3)& 0.11 (5) \\
kinimpact& M & H & H & H & H & H & H\\[0.3cm]
FNfn10 & $\beta_1$  & $\beta_2$  & $\beta_3$ & $\beta_4$ & $\beta_5$ & $\beta_6$ & $\beta_7$\\
$\bar{\Phi}_{\text{exp}}$ & 0.3 (3) & $-0.16$ (2) & 0.55 (3) & 0.35 (2) & 0.29 (2)& 0.44 (4)& 0.73 (1) \\
kinimpact& H & H & H & H & H & H & H\\[0.3cm]
Titin & $\beta_1$  & $\beta_2$  & $\beta_3$ & $\beta_4$ & $\beta_5$ & $\beta_6$ & $\beta_7$\\
$\bar{\Phi}_{\text{exp}}$ & 0.09 (3) & 0.53 (4) & 0.51 (2) & 0.54 (2) & 0.66 (3)& 0.66 (3)& 0.07 (3) \\
kinimpact& M & H & H & H & H & H & M\\[0.3cm]
CspB & $\beta_1$  & $\beta_2$  & $\beta_3$ & G  & $\beta_4$ & $\beta_5$ & \\
$\bar{\Phi}_{\text{exp}}$ & 0.64 (6) & 0.27 (4) & 0.75 (1) & $-0.06$ (2) & 0.16 (2) & 0.12 (2)&  \\
kinimpact& H & H & H & L & H & H &\\[0.3cm]
L23 & $\beta_1$  & $\alpha_1$ & $\beta_2$  & $\alpha_2$ & $\beta_3$ & $\beta_4$ & $\alpha_3$\\
$\bar{\Phi}_{\text{exp}}$ & 0.08 (1) & 0.03 (1) & 0.20 (2) & 0.34 (2) & 0.10 (3) & 0.29 (3) & 0.02 (1)\\
kinimpact& L & L & M & H & H & H & L\\[0.3cm]
ACBP & $\alpha_1$  & $\alpha_2$ & $\alpha_3$  & $\alpha_4$ & & &\\
$\bar{\Phi}_{\text{exp}}$ & 0.34 (4) & -0.19 (9) & 0.57 (2) & 0.21 (6) &  &  &\\
kinimpact& M & H & H & M &  &  & 
\end{tabular}}

\end{table}

\clearpage

Caption Table 2:\\[0.1cm]
The average $\Phi$-values have been calculated from data published in the following articles: CI2\cite{itzhaki95}, protein L \cite{kim00}, protein G \cite{mccallister00}, src SH3 \cite{riddle99},
$\alpha$-spectrin SH3 \cite{martinez99}, Sso7d \cite{guerois00}, ADA2h \cite{villegas98}, U1A \cite{ternstrom99}, S6 \cite{otzen02}, TNfn3 \cite{hamill00}, FNfn10 \cite{cota01}, Titin \cite{fowler01}, CspB \cite{garcia04}, L23 \cite{hedberg04}, ACBP \cite{kragelund99}. The number in brackets behind an average $\Phi$-value indicates the number of residues in the secondary element for which $\Phi$-values have been measured. Averages taken from many $\Phi$-values are more reliable. The kinetic impact of the secondary elements is derived from the results shown in Table 1 and can attain the values low (L), medium (M), or high (H).

\clearpage

%

\clearpage 
%

\begin{thebibliography}{99}
\bibitem{merlo} Merlo C, Dill KA, Weikl TR. $\Phi$-values in protein folding kinetics have energetic and structural components. Submitted
\end{thebibliography}
\end{document}